\documentclass[conference]{IEEEtran}
\IEEEoverridecommandlockouts
\usepackage{cite}
\usepackage{float}
\usepackage{amsmath,amssymb,amsfonts}
\usepackage{algorithmic}
\usepackage{graphicx}
\usepackage{textcomp}
\usepackage{xcolor}
\usepackage{pgfplots}
\usepgfplotslibrary{groupplots}
\def\BibTeX{{\rm B\kern-.05em{\sc i\kern-.025em b}\kern-.08em
    T\kern-.1667em\lower.7ex\hbox{E}\kern-.125emX}}
\usepackage[dvipsnames]{xcolor}

\usepackage{subcaption}
\usepackage{graphicx}
\usepackage{url}
\usepackage{comment}
\usepackage{glossaries}
\usepackage[colorinlistoftodos]{todonotes}
     
\usepackage{booktabs}     
\usepackage{tabularx}     
\usepackage{caption}      
\usepackage{enumitem}     

\usepackage{amssymb}
\usepackage{pifont} 

\newacronym{5g}{5G}{5$^\text{th}$ Generation}
\newacronym{6g}{6G}{6$^\text{th}$ Generation}
\newacronym{3gpp}{3GPP}{3$^\text{rd}$ Generation Partnership Project}
\newacronym{ai}{AI}{Artificial Intelligence}
\newacronym{aiaas}{AIaaS}{Artificial Intelligence as a Service}
\newacronym{aiml}{AI/ML}{Artificial Intelligence and Machine Learning}
\newacronym{api}{API}{application programming interface}
\newacronym{esfri}{ESFRI}{European Strategy Forum on Research Infrastructures}
\newacronym{kpi}{KPI}{Key Performance Indicator}
\newacronym{nwdaf}{NWDAF}{Network Data Analytics Function}
\newacronym{ran}{RAN}{Radio Access Network}
\newacronym{cn}{CN}{Core Network}
\newacronym{slices}{SLICES}{Scientific Large-scale Infrastructure for Computing/Communication Experimental Studies}
\newacronym{ue}{UE}{User Equipment}
\newacronym{}{}{}
\newacronym{aoa}{AoA}{angle of arrival}
\newacronym{ap}{AP}{access point}
\newacronym{awv}{AWV}{antenna weight vector}
\newacronym{bpsk}{BPSK}{binary phase shift keying}
\newacronym{cots}{COTS}{commercial off-the-shelf}
\newacronym{csi}{CSI}{channel state information}
\newacronym{cnn}{CNN}{convolutional neural network}
\newacronym{dmg}{DMG}{directional multi-gigabit}
\newacronym{edmg}{EDMG}{enhanced directional multi-gigabit}
\newacronym{fl}{FL}{federated learning}
\newacronym{ftm}{FTM}{fine timing measurement}
\newacronym{fpbt}{BT}{beamforming training}
\newacronym{fpga}{FPGA}{field programmable gate array}
\newacronym{gnb}{gNB}{next Generation Node B}
\newacronym{isac}{ISAC}{integrated sensing and communications}
\newacronym{iot}{IoT}{internet of things}
\newacronym{jcas}{JCAS}{joint communications and sensing}
\newacronym{los}{LOS}{line of sight}
\newacronym{mac}{MAC}{media access control}
\newacronym{mimo}{MIMO}{multiple-in multiple-out}
\newacronym{ml}{ML}{Machine Learning}
\newacronym{mmwave}{mmWave}{millimeter-wave}
\newacronym{nlos}{NLOS}{non line of sight}
\newacronym{ofdm}{OFDM}{orthogonal frequency-division multiplexing}
\newacronym{phy}{PHY}{physical layer}
\newacronym{psta}{P-STA}{passive station}
\newacronym{poc}{PoC}{Proof-of-Concept}
\newacronym{ppdu}{PPDU}{physical layer protocol data unit}
\newacronym{pdp}{PDP}{power delay profile}
\newacronym{qos}{QoS}{Quality-of-Service}
\newacronym{qpsk}{QPSK}{quadrature phase shift keying}
\newacronym{rssi}{RSSI}{received signal strength indicator}
\newacronym{sgd}{SGD}{stochastic gradient descent}
\newacronym{slam}{SLAM}{simultaneous localization and mapping}
\newacronym{sta}{STA}{station}
\newacronym{snr}{SNR}{signal noise ratio}
\newacronym{sdr}{SDR}{software defined radio}
\newacronym{tof}{ToF}{time of flight}
\newacronym{trn}{TRN}{training}
\newacronym{usrp}{USRP}{universal software radio peripheral}
\newacronym{ura}{URA}{uniform rectangular array}
\newacronym{waveslam}{waveSLAM}{mmWave-based simultaneous localization and mapping}
\newacronym{wlan}{WLANs}{wireless local area networks}
\newacronym{rsrp}{RSRP}{Reference Signal Received Power}
\newacronym{rsrq}{RSRQ}{Reference Signal Received Quality}
\newacronym{gan}{GAN}{Generative Adversarial Network}
\newcommand{\DeclareAuthor}[3]{%
  \expandafter\newcommand\csname #3\endcsname[1]{\textcolor{#2}{\textbf{#1:} ##1}}%
  \expandafter\newcommand\csname #3text\endcsname[1]{\textcolor{#2}{##1}}%
  \expandafter\newcommand\csname #3del\endcsname[1]{\textcolor{#2!70}{\sout{##1}}}%
}

\usepackage[normalem]{ulem} 

\DeclareAuthor{Pablo}{blue}{pablo}
\DeclareAuthor{David}{red}{david}
\begin{document}

\title{
NextSense: A Semi-Synthetic Sensing Data generation Platform\\

\thanks{This work has been funded by the European Commission Horizon
Europe SNS JU MultiX (GA 101192521).

©2026 IEEE. THIS WORK HAS BEEN ACCEPTED IN IEEE INFOCOM ISAC-FutureG Workshop. Personal use of this material is permitted. Permission from IEEE must be obtained for all other uses, in any current or future media, including reprinting/republishing this material for advertising or promotional purposes, creating new or redistribution to servers or lists, or reuse of any copyrighted component of this work in other work}
}

\author{
    \IEEEauthorblockN{David Rico Menéndez}
    \IEEEauthorblockA{\textit{Telematics Department} \\
    \textit{University Carlos III Madrid}\\
    Madrid, Spain \\
    drico@pa.uc3m.es}
    \and
    \IEEEauthorblockN{Pablo Picazo-Martinez}
    \IEEEauthorblockA{\textit{Telematics Department} \\
    \textit{University Carlos III Madrid}\\
    Madrid, Spain \\
    papicazo@pa.uc3m.es}
    \and
    \IEEEauthorblockN{Antonio de la Oliva}
    \IEEEauthorblockA{\textit{Telematics Department} \\
    \textit{University Carlos III Madrid}\\
    Madrid, Spain \\
    aoliva@it.uc3m.es}
    \and
    \IEEEauthorblockN{Carlos Jesús Bernardos}
    \IEEEauthorblockA{\textit{Telematics Department} \\
    \textit{University Carlos III Madrid}\\
    Madrid, Spain \\
    cjbc@it.uc3m.es}
    \and
    \IEEEauthorblockN{Chathura Sarathchandra}
    \IEEEauthorblockA{\textit{InterDigital Europe}\\
    London, UK \\
    Chathura.Sarathchandra@interdigital.com}
    \and
    \IEEEauthorblockN{Alain Mourad}
    \IEEEauthorblockA{\textit{InterDigital Europe}\\
    London, UK \\
    Alain.Mourad@interdigital.com }
    
}

\maketitle

\begin{abstract}
Emerging integrated sensing and communication (ISAC) applications require large volumes of data, but collecting such datasets in real networks is costly, time consuming, and often infeasible due to limited access to low level measurements. In this paper we present NextSense, an open and modular semi-synthetic data generation platform that consists of a 5G stack, a channel emulator, and an UE emulator. The platform allows users full customization on radio configuration, channel and mobility, and traffic profiles through an API and GUI, and produces multi-perspective outputs that combine symbol-level IQ samples, protocol traces, and key performance indicators across UE, RAN, and CN. This paper describes the NextSense's architecture, and validates its ability to act as a faithful proxy for real measurements in sensing use cases.
\end{abstract}

\begin{IEEEkeywords}
5G, 6G, Integrated Sensing and Communication (ISAC), semi-synthetic data, channel emulation, testbeds, open platforms, FutureG
\end{IEEEkeywords}



\section{Introduction}

\gls{ai} and \gls{ml} systems continue to demand ever-larger volumes of data as models grow deeper, more complex, and more data-hungry. In particular, next-generation sensing platforms, envisioned for 6G and beyond, require rich, high-dimensional datasets (raw IQ samples, fine-grained channel state information, mobility traces, propagation logs) to learn robust models for environment inference, localization, mapping, and simultaneous communications and sensing \cite{survey1}. However, this demand has a number of practical impediments.

\begin{figure}[t]
    \centering
\includegraphics[width=0.95\columnwidth]{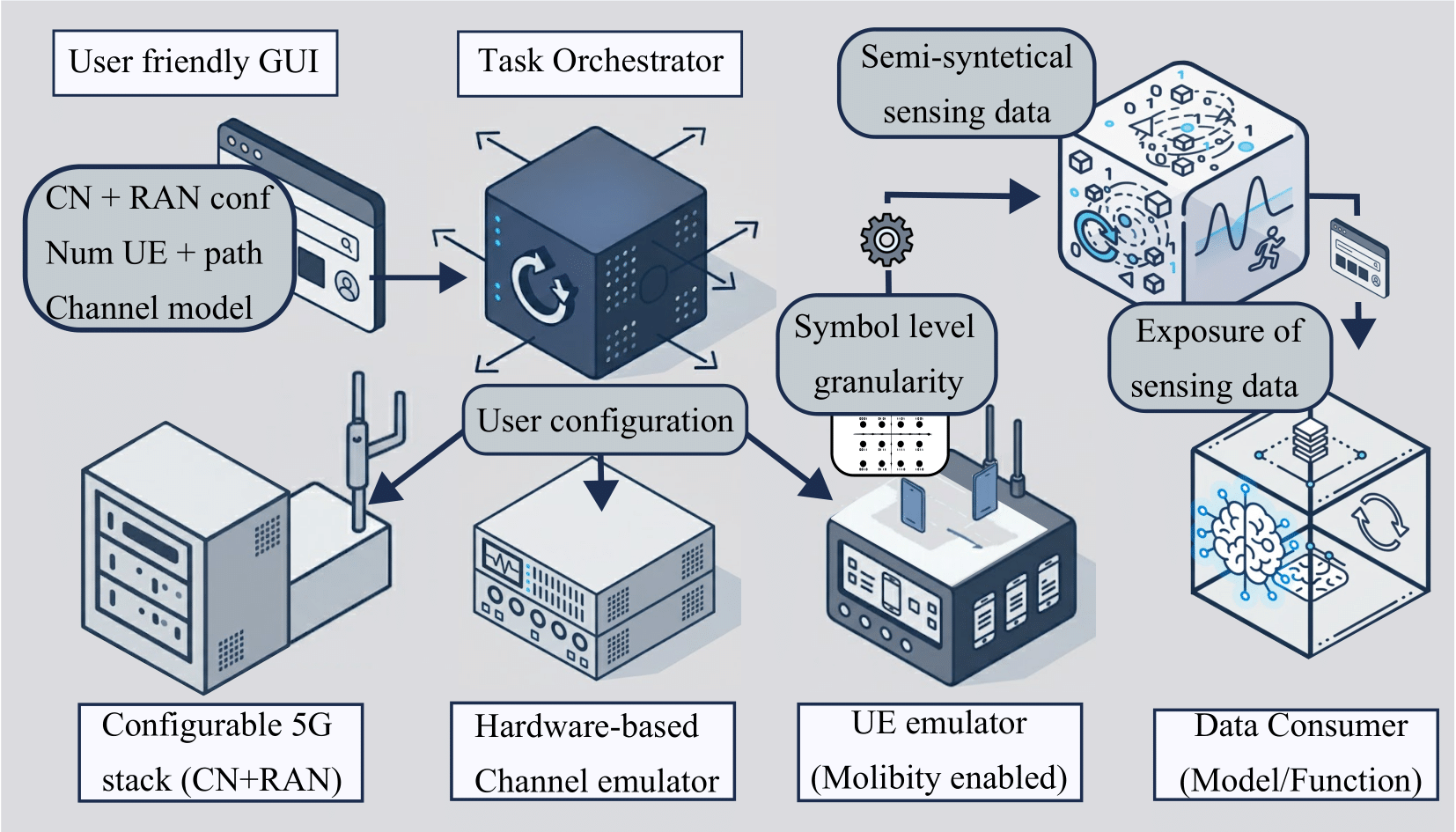}
    \caption{Schematic of the NextSense open testbed. The platform orchestrates an open-source 5G stack, a hardware channel emulator, and a UE emulator to automatically generate configurable, semi-synthetic multi-layer sensing datasets.}
    \label{fig:intro}
    \vspace{-3ex}
\end{figure}

First, mobile network operators are rarely willing or able to grant external researchers access to raw L1 data from commercial deployments, mainly due to privacy, regulatory, business-model, and competitive-sensitivity reasons. Likewise, equipment vendors or system integrators rarely expose the full-granularity of IQ sample streams or propagation characteristics required for detailed training in ML models \cite{mobileop}. As a result, acquiring sufficiently large, labeled, high-fidelity datasets from real networks is often infeasible, slowing down progress towards data-centric sensing and AI-enabled 6G platforms.

To overcome this barrier, we propose a modular semi-synthetic data platform tailored for future 6G sensing research. The platform is built upon an open-source 5G \gls{cn}, configurable radio front-ends (with user-defined mobility within a cell), a customizable hardware channel emulator and a \gls{ue} emulator. The user can instantiate multiple \gls{ue}s with arbitrary trajectories, define propagation scenarios, and capture layer-1 traces: from raw IQ samples to signal-level metadata (\gls{rsrp}, \gls{rsrq}, \gls{snr}, delay spreads), and propagation logs, see Fig. \ref{fig:intro}.

This enables repeatable,  labeled high-granularity synthetic datasets for \gls{ml} training and benchmarking of sensing-aided communications and localization algorithms. The platform bridges the gap between commercial-grade network data scarcity and the needs of \gls{ai} \gls{ml} for next-generation sensing in cellular networks.

The generation of synthetic-data from \gls{5g} and beyond datasets has been explored with \gls{gan}s \cite{syntetic1,syntetic2} or with simulations \cite{syntetic3}. Hardware emulations are emerging as an intermediate solution between pure models and simulations, and real data, offering considerable potential that has already been explored in fields such as security \cite{emulationsemi}. 
However, to the best of our knowledge, there exist no open works that generate semi-synthetic data of a \gls{5g} stack allowing full user customization of parameters such as cell configuration, customized channel models, or \gls{ue} movement patterns. While the use of hardware channel emulators is standard in industry testing, these setups are typically closed, proprietary, and manual. The core methodological innovation of NextSense lies in providing a fully automated, API-orchestrated, and open-source-based framework that democratizes access to granular, layer-1 sensing data (e.g., raw IQ samples) typically locked inside commercial vendor equipment. Thus, we fill the gap in the literature by offering the first open testbed to generate semi-synthetic \gls{5g} sensing data for B5G.

We believe that this can help researchers to enhance the performance of their models by having access to large amounts of data, generable using a friendly interface and following an open, fully customizable 5G stack.
 
The remainder of the paper is organized as follows. Section 2 provides an overview of the system. Section 3 describes the interfaces and configuration options of the platform. Section 4 explains how the output of the system is structured and demonstrates an example execution of the platform to generate sensing data. Section 5 presents baseline comparisons between semi-synthetic data produced by our system and real-world data. Finally, Section 6 concludes the paper.

\section{System Overview and Orchestration}

The system is composed of three main components: an OAI 5G stack \cite{oai}, an Amarisoft UE Emulator, and a Keysight PROPSIM F64 channel emulator. Together, they form a fully controllable end-to-end 5G emulation testbed. Each component exposes a standardized interface to the Orchestrator, which manages experiment deployment, synchronization, and runtime control. Through this interface, configurations can be automatically loaded, started, and monitored across all modules, enabling seamless execution of complex network and sensing experiments.
\subsection{OAI 5G Core}
The OAI Core provides a fully containerized 5G Core Network compliant with 3GPP Release 17. It includes the main network functions such as AMF, SMF, UPF, and NRF. The orchestrator automatically deploys these functions using Docker Compose generated runtime based on the experiment configuration. The core can interoperate with both simulated and physical gNBs, supporting data-plane measurements and NWDAF-based analytics.

\subsection{OAI gnB}
The OAI gNB acts as the radio access node in the deployment. Connects to the CN via the NG interface and operates in FR1, depending on the experiment configuration.

The orchestrator manages gNB configuration dynamically, setting parameters such as frequency band, transmission power, and SSB periodicity, among others. Integration scripts ensure compatibility between the gNB and UE, enabling automated attachment, registration, and data-plane establishment.

\subsection{Channel hardware emulator}
The Keysight PROPSIM F64 channel emulator provides a controlled and repeatable radio environment for testing the 5G end-to-end setup. It emulates multipath fading, Doppler, and mobility effects across multiple frequency bands and MIMO configurations.

The orchestrator connects to the PROPSIM via its remote control API, automatically loading the selected channel scenario (e.g., 3GPP TR 38.901 urban, indoor, or custom profiles) and synchronizing playback with the gNB and UE startup.

Key features include up to 400 MHz bandwidth, 64 RF channels for massive MIMO, and real-time scenario control through SCPI or REST interfaces. External clock and GPS synchronization ensure alignment with the rest of the testbed, enabling accurate and reproducible performance evaluation.

\subsection{UE Emulator}
The Amarisoft UE Emulator provides a flexible and software-defined implementation of the 5G user equipment. It can emulate multiple UEs simultaneously, each with configurable radio parameters and traffic profiles. It also contains certain channel emulation capabilities on its own, that can be used toggled on and off as user demands.

Through the orchestrator, the UE emulator is launched and connected to the OAI gNB to initiate the UE's registration procedure. During runtime, it can generate user traffic or sensing data streams as defined in the experiment configuration. Bidirectional control between the orchestrator and the Amarisoft UE ensures consistent experiment execution and synchronization with the other components.

\begin{figure*}[t]
    \centering
    \includegraphics[width=\textwidth]{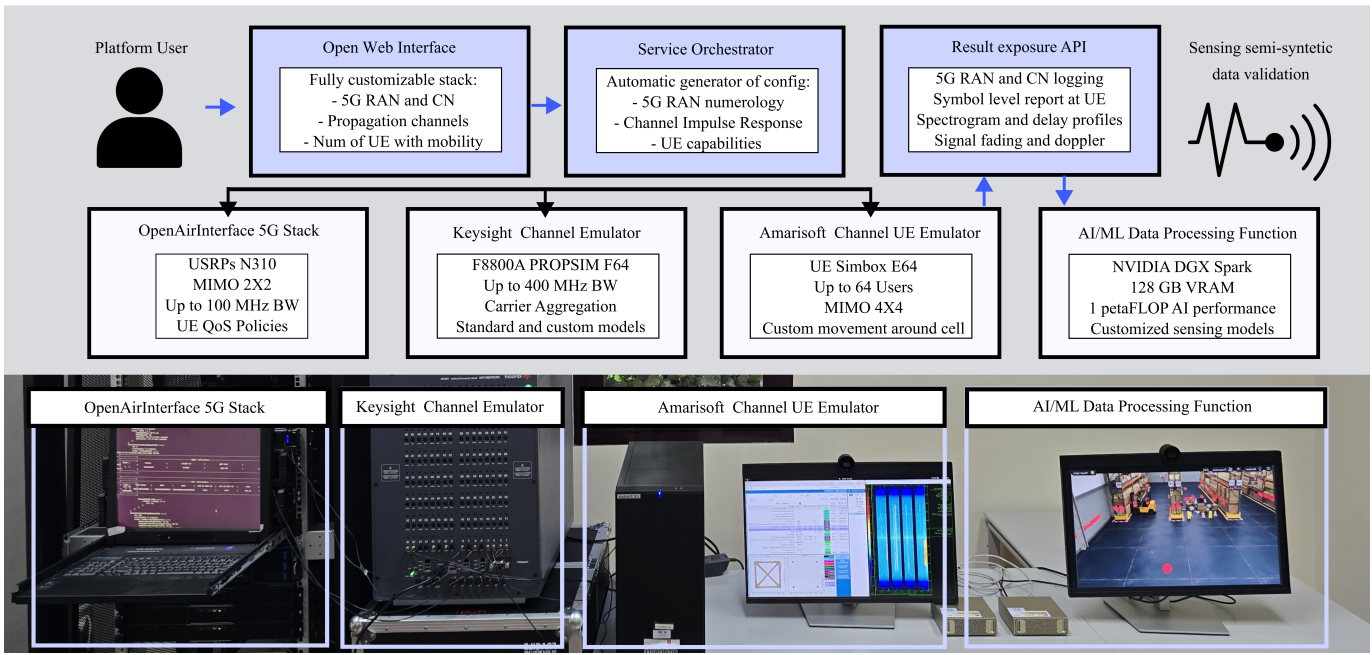}
    \caption{High-level architecture and physical realization of the NextSense semi-synthetic sensing platform.}
    \vspace{-3mm}
    \label{fig:architecture}
\end{figure*}

\section{Input Scenario Parametrization: Channel, Traffic, Mobility}

The strength of our platform lies in its ability to define high-fidelity, fully automated, and repeatable experimentation scenarios. As illustrated in Fig.~\ref{fig:architecture}, the task orchestrator ingests a user configuration that comprehensively parameterizes the entire 5G stack (CN+RAN), the radio environment, and the UE's behavior. While the platform's GUI offers a simplified selection of the most relevant parameters for ease of use, maximum granularity and full parameter control are achieved via the API. This input parametrization is crucial as it defines the feature space of the semisynthetic sensing data to be generated, and it is summarized in the \emph{Input Parameters} column of Table~\ref{tab:io_parameters}.

The scenario parameters are grouped into three fundamental pillars: radio configuration, channel and mobility emulation, and traffic profiles.

\begin{enumerate}
    \item Radio Network Configuration (gNB and UE)
    
The platform provides granular control over the physical and media access layers, allowing the user to configure the gNB (based on OAI) and the emulated UEs (Amarisoft). This includes key signal parameters such as the operating frequency, bandwidth (e.g., 1.4 to 20 MHz in LTE; up to 100 MHz in NR), numerology (with \texttt{subcarrier\_spacing} from 15 to 120 kHz in NR), and transmission and reception powers. These options correspond to the \textit{Radio \& gNB Configuration} group of Table~\ref{tab:io_parameters}.

\item Channel Emulation and Mobility

The second pillar of parametrization is the radio environment, simulated using a combination of the Amarisoft channel emulator and the Keysight PROPSIM hardware emulator. This duality allows for everything from rapid simulations to high-fidelity emulations.

The platform supports a wide catalog of standard 3GPP channel models, essential for validation. This includes classic NR Tapped Delay Line (TDL) profiles (e.g., \texttt{tdla30}, \texttt{tdlb100}, \texttt{tdlc300}) as well as dense urban, vehicular (V2X), or non-terrestrial network (NTN) scenarios. The user can specify every key channel parameter such as \texttt{delay\_spread} (in ns), Doppler frequency (\texttt{freq\_doppler}), and MIMO correlation (low, medium, high), as well as path loss coefficients (A, B) for custom models. These parameters appear in the \textit{Channel Emulation} block of Table~\ref{tab:io_parameters}.

A distinctive capability of our architecture is the integration with the Keysight channel emulator, which allows the orchestrator to load custom and dynamic scenarios. This includes the ability to import complex channel profiles created in external tools like MATLAB, opening the door to simulating very specific environments. Together with the UE mobility and behavior options in Table~\ref{tab:io_parameters}, this enables precise control of how users move and interact with the emulated radio environment.
\end{enumerate}


\begin{table}[htbp]
  \centering
  \caption{Platform Input/Output Parameter Space}
  \label{tab:io_parameters}

  \begin{tabularx}{\columnwidth}{X || X}
    \toprule
    \textbf{Input Parameters} & \textbf{Output Data \& Logs} \\
    \midrule

    \vtop{
      \textit{\textbf{Radio \& gNB Configuration}}
      \begin{itemize}[leftmargin=*, noitemsep, topsep=0pt]
        \item Number of cells * +
        \item Antenna position + \#
        \item Antenna type + \#
        \item NR Frequency * +
        \item Bandwidth * +
        \item Subcarrier Spacing * +
        \item TX Power * +
        \item Nº of DL/UL Antennae * +
        \item Rx to Tx latency * +
        \item Max MCS * +
      \end{itemize}

      \textit{\textbf{Channel Emulation}}
      \begin{itemize}[leftmargin=*, noitemsep, topsep=0pt]
        \item UE type +
        \item 3GPP TDL Models + \#
        \item 3GPP Fading Profiles + \#
        \item Delay Spread + \#
        \item Doppler Frequency + \#
        \item MIMO Correlation + \#
        \item Path Loss Coeffs. + \#
        \item Noise Spectral Density + \#
        \item Custom MATLAB Scenario Import \#
      \end{itemize}

      \textit{\textbf{UE Mobility \& Behavior}}
      \begin{itemize}[leftmargin=*, noitemsep, topsep=0pt]
        \item Initial Position (X,Y,Z) +
        \item Speed, Direction, Elevation + 
        \item Mobility Logic/Area +
      \end{itemize}

      \textit{\textbf{Traffic \& Data Collection}}
      \begin{itemize}[leftmargin=*, noitemsep, topsep=0pt]
        \item Traffic Profile +
        \item Log Verbosity Control * +
      \end{itemize}
    }

    &

    \vtop{
      \textit{\textbf{UE Sensing Data}}
      \begin{itemize}[leftmargin=*, noitemsep, topsep=0pt]
        \item Symbol-Level IQ Samples +
        \item UE Mobility Traces (X,Y,Z) +
        \item UE-measured KPIs (RSRP, RSRQ, etc.) +
      \end{itemize}

      \textit{\textbf{UE Protocol Logs}}
      \begin{itemize}[leftmargin=*, noitemsep, topsep=0pt]
        \item PHY / MAC / RLC / PDCP Logs +
        \item RRC / NAS / IP Layer Logs +
      \end{itemize}

      \textit{\textbf{gNB Protocol \& Signaling Logs}}
      \begin{itemize}[leftmargin=*, noitemsep, topsep=0pt]
        \item RRC/PDCP/RLC Setup *
        \item RRC Reconfiguration *
        \item E1AP/NGAP PDU Session *
      \end{itemize}

      \textit{\textbf{gNB Per-UE KPI Logs}}
      \begin{itemize}[leftmargin=*, noitemsep, topsep=0pt]
        \item MAC Scheduler Timestamps *
        \item RSRP *
        \item DL/UL BLER \& MCS *
        \item DL/UL Rounds (Retransmissions) *
        \item Uplink SNR *
        \item MAC-level TX/RX Bytes *
      \end{itemize}

      \textit{\textbf{Core Context Logs}}
      \begin{itemize}[leftmargin=*, noitemsep, topsep=0pt]
        \item AMF Logs *
        \item Other Core NF Logs *
      \end{itemize}

      \textit{\textbf{Processed KPIs}}
      \begin{itemize}[leftmargin=*, noitemsep, topsep=0pt]
        \item Packet Capture * +
        \item TA (Timing Advance) *
        \item Throughput (BRATE) * +
      \end{itemize}
    }

    \\
    \bottomrule
  \end{tabularx}

  \parbox{\columnwidth}{
  \vspace{5pt} 
  \footnotesize
  \textbf{Legend:} \\
  \textbf{*} = OAI (gNB/Core) \quad
  \textbf{+} = Amarisoft (UE/Em) \quad
  \textbf{\#} = Keysight (Channel Emu)
  }
  \vspace{-3mm}

\end{table}

\section{Data Collection and Example Run}

Following the execution of a parametrized scenario, the platform's orchestrator collects, aggregates, and stores a comprehensive set of logs from every component in the end-to-end chain. This data serves as the ''ground truth'' output, forming the basis for the semisynthetic dataset and populating the \emph{Output Data \& Logs} column of Table~\ref{tab:io_parameters}. The data collection is strategically divided between the UE, the RAN, and the Core Network, providing a holistic view of the experiment.

\begin{enumerate}
    \item UE-centric Data (Amarisoft)
    
The most granular and valuable data for sensing originates from the emulated UEs. Our platform leverages the advanced logging capabilities of the Amarisoft stack, which provides full visibility into the UE's protocol stack. As specified by the user, logs are captured across all layers (PHY, MAC, RLC, PDCP, RRC, NAS, IP).

A key feature of our architecture is the ability to define the \texttt{log\_options} verbosity for each layer. This fine-grained control is paramount for sensing applications. The platform enables up to capture of symbol-level IQ samples. This can be achieved in two different ways, by capturing IQ samples of desired signals, such as PDSCH, SSB, etc. or by capturing the whole RX/TX channel. This last one provides a raw, unprocessed view of the radio channel as experienced by the UE, which is ideal for training advanced AI/ML models for localization, environmental inference, and passive radar. While the full, multi-layer debug logs can be exceptionally large they are captured in their entirety for post-processing and filtering, as summarized under \textit{UE Sensing Data} and \textit{UE Protocol Logs} in Table~\ref{tab:io_parameters}.

\item 5G RAN and Core Network Logs

To complement the UE-centric data, the platform also captures execution logs from the 5G OAI-based infrastructure stack, providing essential context for the experiment.

\begin{itemize}
    \item gNodeB Logs: The standard output and execution logs from the OAI gNB are captured. This data details radio link status, RRC-level decisions, resource allocation, MCS, Uplink and Downlink info, and PHY/MAC layer procedures from the network's perspective.

    \item CN Logs: We capture logs from the OAI 5G Core, by default from the AMF (Access and Mobility Management Function) and configurable to any other desired network function. This provides crucial context on NAS signaling, UE registration/deregistration procedures, PDU session management, and mobility events.
\end{itemize}

This data collection strategy provides a synchronized, multi-perspective view of each experiment. The UE-side IQ and RRC logs (detailing what the UE sensed) are contextualized by the gNB and CN logs (detailing why the network behaved as it did). As depicted in Fig.~\ref{fig:architecture}, this aggregated data is then forwarded to the AI Service block (NVIDIA DGX Spark), where it can be processed, labeled, and prepared for consumption by sensing model developers.
\end{enumerate}





\subsection{End to End Example Run}
\begin{figure}[t]
    \centering
    \includegraphics[width=0.95\columnwidth]{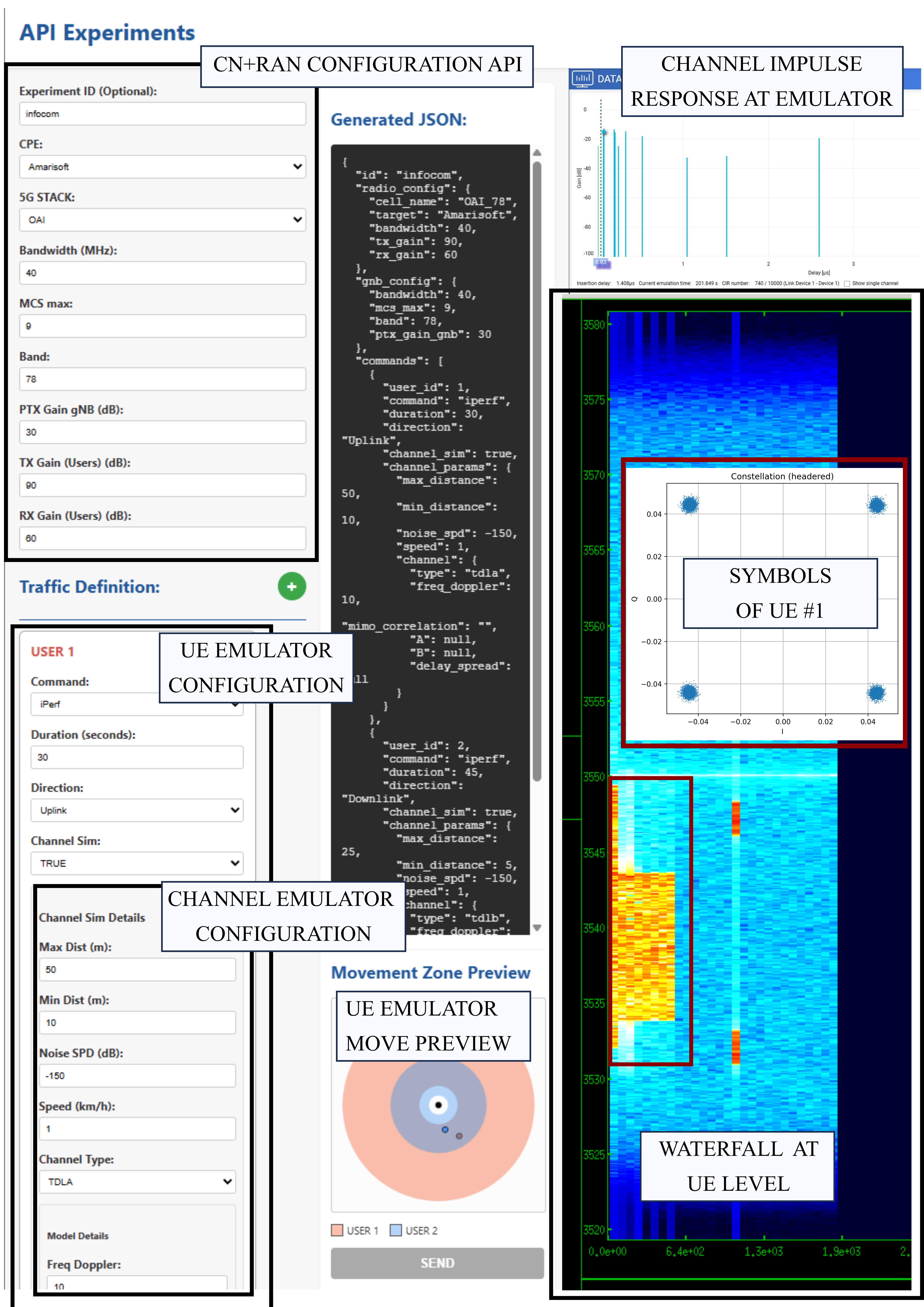}
    \caption{End-to-end workflow of a NextSense experiment. The user defines the API parameters (left), the orchestrator automatically applies the channel impulse response (top right), and the platform exposes synchronized multi-layer outputs, including symbol-level UE sensing data (bottom right).}
    \label{fig:run}
    \vspace{-3ex}
\end{figure}

Fig.~\ref{fig:run} illustrates a complete example of how a researcher interacts with the platform, from IN parametrization to OUT sensing data. On the left side, the \emph{API Experiments} panel allows the user to specify all the scenario parameters discussed above, including CN+RAN configuration, UE emulator options, and channel emulation settings. These fields correspond directly to the input space in Table~\ref{tab:io_parameters}. Once the form is filled, the orchestrator generates the corresponding JSON specification, shown in the \emph{Generated JSON} panel. Advanced users can export or edit this JSON and invoke the API directly using tools such as Postman, bypassing the GUI if desired.

Below the JSON panel, the \emph{Movement Zone Preview} provides a real-time visualization of the UE emulator mobility configuration before launching the experiment. This preview allows the user to verify the initial positions, trajectories, and interaction area of each UE, ensuring that the selected mobility profiles match the intended scenario.

After the configuration is validated and the experiment is started, the orchestrator configures the channel emulator and the radio stack according to the JSON specification. The top-right part of Fig.~\ref{fig:run} shows an example channel impulse response at the emulator, confirming that the selected emulated channel and delay spread are correctly applied. In parallel, the UE emulator connects to the configured CN+RAN and starts to transmit and receive according to the chosen traffic profile.

The bottom-right part of Fig.~\ref{fig:run} illustrates two representative sensing outputs collected at the UE side. The large panel shows a time–frequency waterfall at the UE, obtained from symbol-level IQ capture, while the inset displays the constellation of a selected reference signal for UE~\#1. These outputs are particular instances of the \textit{UE Sensing Data} in Table~\ref{tab:io_parameters}. Together with the protocol and KPI logs collected from the gNB and the CN, this example run demonstrates how a single JSON experiment description is transformed into a rich, multi-perspective semisynthetic dataset ready for downstream AI/ML sensing tasks.

\section{Validation in a real Use Case Scenario}

\begin{figure*}[t]
    \centering
    \includegraphics[width=\textwidth]{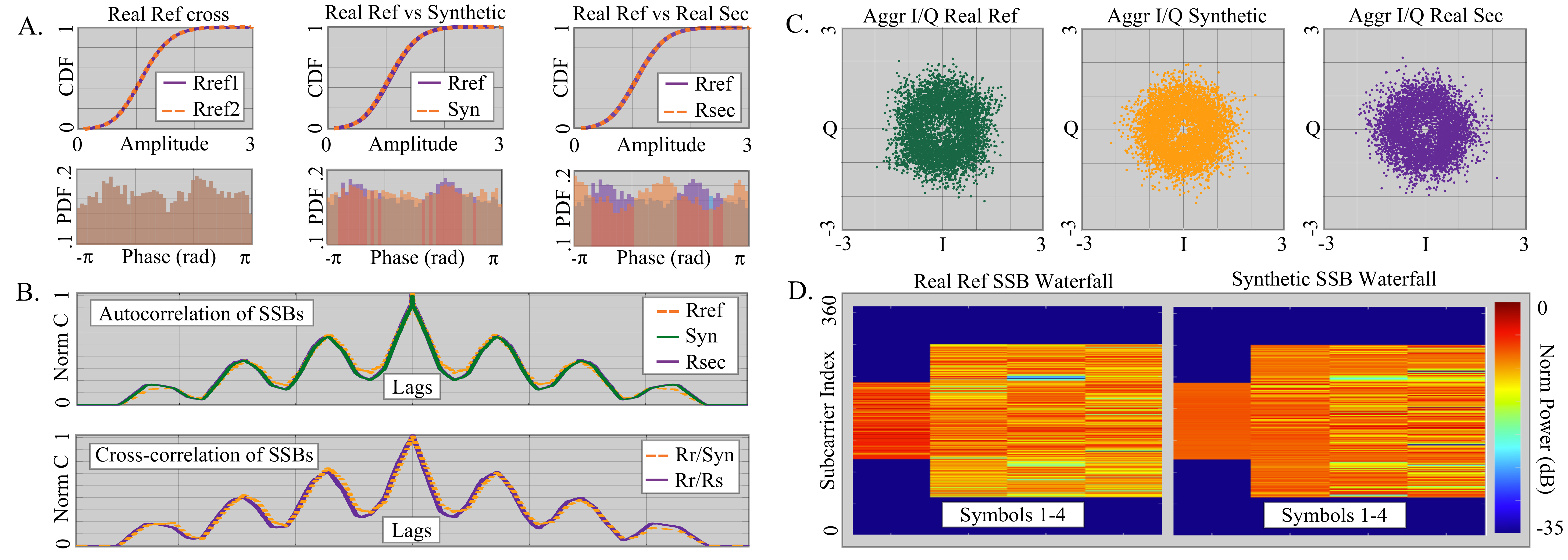}
    \caption{Experimental results on synthetic data generation: A) Distributions on amplitude and phase (left) Real ref split, (middle) Real ref vs Synthetic, and (right) Real ref vs sec; B) Temporal autocorrelation (top) and cross-correlation (bottom); C) Aggregated I/Q samples for (left) Real ref, (middle) Synthetic, (right) Real sec; D) Waterfall SSB Subcarrier Power of (left) Synthetic and (right) Real ref.}
    \label{fig:validation}
    \vspace{-3mm}
\end{figure*}

To demonstrate NextSense's capability to generate high–fidelity synthetic data suitable for training AI/ML sensing models, we conducted a comparative validation against a real–world reference dataset.

\subsection{Reference Scenario and Methodology}

The validation scenario focuses on an Indoor Presence Detection use case. We captured real–world channel traces (CSI/SSB), a well known signal that can be used in passive sensing \cite{ssb}, in a controlled indoor environment under three distinct conditions: Empty Room, Human Presence, and Mobile Robot Presence. For each condition, the dataset consists of a temporal sequence of SSB snapshots captured continuously over time. From this dataset we extract two real reference ensembles: a reference SSB set and a second set recorded after moving the antennae, which changes the phase of the channel while keeping its average power essentially unchanged.

%

%
To build the emulated sensing channel used in the validation, we follow a data-driven procedure that maps real measurements into a tap-based channel impulse response compatible with the PROPSIM emulator. For each snapshot, we treat the SSB as a known reference signal $X(k)$ and estimate the frequency-domain channel on the SSB subcarriers as
\begin{equation}
    \hat{H}(k,n) = \frac{Y(k,n)}{X(k)},
\end{equation}
where $Y(k,n)$ is the received SSB at subcarrier $k$ and snapshot index $n$. We then obtain a discrete-time channel impulse response by applying an inverse DFT across subcarriers,
\begin{equation}
    \hat{h}(\ell,n) = \mathrm{IDFT}_k\{\hat{H}(k,n)\},
\end{equation}
and compute the empirical power delay profile (per tap $\ell$) $P(\ell) = \mathbb{E}_n[|\hat{h}(\ell,n)|^2]$ and its associated delay spread and Doppler profile. From these statistics we select a finite set of dominant taps, each characterized by a delay $\tau_\ell$, average power $P_\ell$ and Doppler $f_{D,\ell}$, and encode them into the PROPSIM scenario description. This way, each sensing scenario is associated with a specific time-varying channel response that reproduces the multipath structure, and fading profile observed in the real measurements. Beyond this data-driven workflow, NextSense also supports arbitrary user-defined channels generated offline (e.g., in MATLAB), which are translated into the same tap-based representation and loaded into the channel emulator.

\textbf{Statistical ensemble approach:} The validation is carried out on the aggregate statistical ensemble of the captured SSBs. Since thermal noise and instantaneous interference are stochastic, a deterministic one–to–one comparison between single real and synthetic snapshots is neither feasible nor meaningful. Instead, our objective is to verify that the NextSense platform generates SSB ensembles whose statistical properties (distribution, spectral shape, and temporal correlation) converge to the same global distribution as the real–world measurements, that is, $X_{\text{real}} \sim X_{\text{syn}}$.

\subsection{Quantitative Statistical Analysis}

We perform a comparative analysis on power–normalised data tensors composed of aggregated SSB snapshots from each domain. We captured two real SSBs (1st \textit{reference} and 2nd \textit{secondary}), each of those with the receiver antennae shifted 90º, and we generated a synthetic SSB with NextSense. Both real and synthetic tensors have dimensions $360 \times 4 \times 100$ (subcarriers, symbols, snapshots).

\textbf{Energy consistency:} The synthetic magnitude variance ($\mathrm{Var}(|X_{\text{syn}}|) = 0.1129$) matches the variability of a secondary real set ($\mathrm{Var}(|X_{\text{real2}}|) = 0.1129$) within a $1\,\%$ relative deviation from the reference ($\mathrm{Var}(|X_{\text{real}}|) = 0.1122$). This confirms that the emulator preserves the channel's energy dynamics within natural capture variability.

\textbf{Kolmogorov–Smirnov (KS) test \cite{kolmosmir}:} Comparisons of real-synthetic ($D = 0.0037, p = 0.629$) and real-real ($D = 0.0038, p = 0.597$) distributions yield $p \gg 0.05$. These nearly identical results indicate no statistical grounds to reject the null hypothesis, placing the synthetic ensemble well within the distributional variability of real measurements.

\textbf{Wasserstein distance (WD) \cite{wass}:} The distance between real and synthetic sets ($\mathrm{WD}_{\text{real,syn}} = 0.00149$) is virtually indistinguishable from the real-to-real variability ($\mathrm{WD}_{\text{real,real2}} = 0.00151$). This demonstrates that the model captures the fine-grained probability density function (PDF) shape as accurately as independent real acquisitions.

\subsection{Spectral and Temporal Domain Validation}
Beyond scalar metrics, Fig.~\ref{fig:validation} shows that real and semi-synthetic SSB ensembles match in the frequency, time, and complex domains. Magnitude CDFs and phase statistics closely overlap, while phase shifts due to antenna displacement mainly affect phase but not energy. Temporal autocorrelation and cross-correlation curves exhibit the same coherence behavior, indicating that fading dynamics are preserved. 
The aggregated sensing-domain I/Q samples (Fig. 4C) show comparable spread, but are intended to validate statistical channel behavior rather than demonstrating communication-level modulation or symbol structure. Meanwhile, the SSB waterfall (Fig. 4D) confirms similar frequency-selective fading patterns.

Overall, NextSense successfully reproduces the statistical properties and distributional variability of the channel, generating sensing signals with high statistical realism.

\subsection{Validation with ML model}
We further assess transferability with a binary sensing task where a mobile robot blocks (or not) the LoS path between UE and gNB. An SVM is trained only on NextSense semi-synthetic data and evaluated on real measurements with matched settings, without fine tuning. The classifier \cite{model} achieves 98.33\% accuracy on real data, indicating a small domain gap and supporting semi-synthetic pre-training before limited real-world adaptation.



\section{Conclusions and Future Work}
We presented NextSense, an open and modular platform to generate semi-synthetic sensing datasets from an end-to-end 5G setup. By orchestrating OAI (CN+gNB), a Keysight PROPSIM channel emulator, and an Amarisoft UE emulator, the platform exposes a rich scenario input space and produces synchronized outputs combining symbol-level IQ, protocol traces, and KPIs across UE, RAN, and CN.

We validated NextSense in an indoor presence-detection setting by reconstructing real SSB-derived channel statistics in the emulator and generating matched semi-synthetic snapshots. Statistical tests (variance, KS, Wasserstein, and temporal correlation) demonstrate high statistical and distributional fidelity, providing a solid baseline for ML models relying on magnitude and basic temporal correlations, though further validation of coherent signal-level structure remains part of our future roadmap.

As future work, we will expand channel and mobility catalogs, broaden radio configurations, and streamline post-processing and labeling for downstream AI workflows.

\bibliographystyle{IEEEtran}

\bibliography{ref.bib}

\begin{thebibliography}{10}
\providecommand{\url}[1]{#1}
\csname url@samestyle\endcsname
\providecommand{\newblock}{\relax}
\providecommand{\bibinfo}[2]{#2}
\providecommand{\BIBentrySTDinterwordspacing}{\spaceskip=0pt\relax}
\providecommand{\BIBentryALTinterwordstretchfactor}{4}
\providecommand{\BIBentryALTinterwordspacing}{\spaceskip=\fontdimen2\font plus
\BIBentryALTinterwordstretchfactor\fontdimen3\font minus \fontdimen4\font\relax}
\providecommand{\BIBforeignlanguage}[2]{{%
\expandafter\ifx\csname l@#1\endcsname\relax
\typeout{** WARNING: IEEEtran.bst: No hyphenation pattern has been}%
\typeout{** loaded for the language `#1'. Using the pattern for}%
\typeout{** the default language instead.}%
\else
\language=\csname l@#1\endcsname
\fi
#2}}
\providecommand{\BIBdecl}{\relax}
\BIBdecl

\bibitem{survey1}
S.~Aldirmaz-Colak \emph{et~al.}, ``A comprehensive review on isac for 6g: Enabling technologies, security, and ai/ml perspectives,'' \emph{IEEE Access}, vol.~13, pp. 97\,152--97\,193, 2025.

\bibitem{mobileop}
M.~Ghahramani, M.~Zhou, and G.~Wang, ``Urban sensing based on mobile phone data: approaches, applications, and challenges,'' \emph{IEEE/CAA Journal of Automatica Sinica}, vol.~7, no.~3, pp. 627--637, 2020.

\bibitem{syntetic1}
C.~Pandey, V.~Tiwari, J.~J. P.~C. Rodrigues, and D.~S. Roy, ``5gt-gan-net: Internet traffic data forecasting with supervised loss based synthetic data over 5g,'' \emph{IEEE Transactions on Mobile Computing}, vol.~23, no.~11, pp. 10\,694--10\,705, 2024.

\bibitem{syntetic2}
H.~Vietz, M.~Hirth, S.~Baum, and M.~Weyrich, ``Synthetic data generation for improving deep learning-based 5g indoor positioning,'' in \emph{2023 IEEE 28th International Conference on Emerging Technologies and Factory Automation (ETFA)}, 2023, pp. 1--7.

\bibitem{syntetic3}
S.~M.~A. Zaidi, M.~Manalastas, H.~Farooq, and A.~Imran, ``Syntheticnet: A 3gpp compliant simulator for ai enabled 5g and beyond,'' \emph{IEEE Access}, vol.~8, pp. 82\,938--82\,950, 2020.

\bibitem{emulationsemi}
S.~Myneni \emph{et~al.}, ``Unraveled — a semi-synthetic dataset for advanced persistent threats,'' \emph{Computer Networks}, vol. 227, p. 109688, 2023.

\bibitem{oai}
S.~Kumar \emph{et~al.}, ``Openairinterface as a platform for 5g-ntn research and experimentation,'' in \emph{2022 IEEE Future Networks World Forum (FNWF)}, 2022, pp. 500--506.

\bibitem{ssb}
K.~Abratkiewicz \emph{et~al.}, ``Ssb-based signal processing for passive radar using a 5g network,'' \emph{IEEE Journal of Selected Topics in Applied Earth Observations and Remote Sensing}, vol.~16, pp. 3469--3484, 2023.

\bibitem{kolmosmir}
F.~J. Massey~Jr, ``The kolmogorov-smirnov test for goodness of fit,'' \emph{Journal of the American statistical Association}, vol.~46, no. 253, pp. 68--78, 1951.

\bibitem{wass}
V.~M. Panaretos and Y.~Zemel, ``Statistical aspects of wasserstein distances,'' \emph{Annual review of statistics and its application}, vol.~6, no.~1, pp. 405--431, 2019.

\bibitem{model}
M.~Pal and P.~M. Mather, ``Support vector machines for classification in remote sensing,'' \emph{International journal of remote sensing}, vol.~26, no.~5, pp. 1007--1011, 2005.

\end{thebibliography}
\end{document}